\documentstyle[aps,prl,preprint,floats,epsfig]{revtex}

\begin{document}

\preprint{\tighten{\vbox{\hbox{CLNS 00/1705 \hfill}
                         \hbox{CLEO 00-26   \hfill} }}}

\tighten
\draft

\title{A Search for Charmless $B \to VV$ Decays}
\author{CLEO Collaboration}
\date{January 12, 2001}

\maketitle
\tighten

\begin{abstract}
We have studied two-body charmless decays of the $B$ meson into the final states $\rho^0 \rho^0$, $K^{*0} \rho^0$, $K^{*0} K^{*0}$,
$K^{*0} \bar{K^{*0}}$, $K^{*+} \rho^0$,
$K^{*+} \bar{K^{*0}}$, and $K^{*+} K^{*-}$ 
using only decay modes with charged daughter particles. Using 9.7 million $B \bar{B}$ pairs collected with the CLEO detector, we place 90\% confidence level upper limits on the branching fractions, $(0.46-7.0)\times 10^{-5}$, depending on final state and
polarization.
\end{abstract}

\clearpage

{
\renewcommand{\thefootnote}{\fnsymbol{footnote}}

\begin{center}
R.~Godang,$^{1}$
G.~Bonvicini,$^{2}$ D.~Cinabro,$^{2}$ M.~Dubrovin,$^{2}$
S.~McGee,$^{2}$ G.~J.~Zhou,$^{2}$
A.~Bornheim,$^{3}$ E.~Lipeles,$^{3}$ S.~P.~Pappas,$^{3}$
M.~Schmidtler,$^{3}$
% ...
\end{center}

\small
\begin{center}
$^{1}${Virginia Polytechnic Institute and State University,
Blacksburg, Virginia 24061}\\
$^{2}${Wayne State University, Detroit, Michigan 48202}\\
$^{3}${California Institute of Technology, Pasadena, California 91125}\\
% ...
\end{center}
\setcounter{footnote}{0}
}
\newpage

In the Standard Model, $\rm CP$ violation is introduced by the complex phase 
in the Cabibbo-Kobayashi-Maskawa quark-mixing matrix. 
The experimental study of CKM phases will
probe the Standard Model description of CP violation.
This may provide a window to new physics. In particular, it
has been suggested\cite{soni} 
that we may construct a relationship between charmless $B\to VV$ decays that
may lead to the extraction of the angle $\alpha$.
Earlier observations of 
rare charmless decay modes at CLEO include
 $B \to K\pi,\pi\pi,\eta K,\rho\pi,\eta^\prime K,\eta K^*$ and
$\omega\pi$ \cite{otherCLEO}. 
It is natural to extend 
our search toward other rare charmless $B$ decays.

In this letter, we present results of searches for $B$ meson decays into the 
vector mesons $\rho^0$, $K^{*0}$ and $K^{*+}$. 
The decays are dominated by the $b \to u$ 
tree-level and $b \to dg$ penguin processes, though other mechanisms may also contribute 
 \cite{theorists}.

The data used in this analysis were collected by the 
CLEO detector\cite{kubota} 
at the Cornell Electron Storage Ring (CESR). 
The data consist of an integrated luminosity of 9.1 fb$^{-1}$ at 
the $\Upsilon(4S)$ resonance, corresponding to $9.7 \times 10^6$ 
$B\bar{B}$ events. To determine
backgrounds due to non-resonant $e^+ e^- \to q\bar{q}$ process,
 we also collected 4.6 fb$^{-1}$ of continuum data at energies 
just below the $\Upsilon(4S)$ resonance.

The CLEO detector has 67 tracking layers and a CsI electromagnetic calorimeter
 that provides efficient $\pi^0$ reconstruction, all operating within a 
$\rm1.5T$ superconducting solenoid. The central tracking system, consisting 
of an inner
6-layer straw tube precision tracker, a 10-layer vertex drift chamber, and a 
51-layer main drift chamber, provides a measurement of momenta of charged 
particles and the vertex position of decaying $K_S$. It also measures the 
specific ionization loss, $dE/dx$, which is used for particle identification. 
The precision tracker 
was replaced by a silicon vertex detector for the latter 65\% of data taking. 
Muons are identified using proportional counters placed at various depths in 
the steel return yoke of the magnet.

$B$ candidates are selected by straightforward criteria based on 
energy-momentum conservation 
and event shape. Simulations of the signal and backgrounds are
used to refine these criteria and to determine their effectiveness.

The $B \to VV$ decays are reconstructed through the decay channels  
$B^0 \to \rho^0 \rho^0$,
$B^0 \to K^{*0} \rho^0$, $B^0 \to K^{*0} K^{*0}$,
$B^0 \to K^{*0} \bar{K^{*0}}$, $B^+ \to K^{*+} \rho^0$,
$B^+ \to K^{*+} \bar{K^{*0}}$, and $B^0 \to K^{*+} K^{*-}$. 
We form $\rho^0$ candidates from $\pi^+ \pi^-$ pairs with an invariant mass within 150 MeV$/c^2$ of 
the nominal $\rho^0$ mass. $K^{*0}/\bar{K^{*0}}/K^{*\pm}$ 
candidates are selected from $K^\pm\pi^\mp/K^0_S \pi^\pm$ 
pairs within 50 MeV$/c^2$ of the nominal $K^*$ mass.

Charged tracks are selected by requiring them  to pass quality criteria and
must be consistent with production from the primary interaction point
(except for pions from $K^0_S$ decays).
The measured specific ionization ($dE/dx$) of charged kaon and pion candidates is required to be within $3.0\sigma$ (standard deviation) of their most probable values. We reject electrons based on $dE/dx$ and the ratio of the track momentum to the associated shower energy in the CsI calorimeter. We reject muons by requiring that the tracks not penetrate the steel absorber past a depth of 3 nuclear interaction lengths. The $K^0_S$ is selected by requiring a decay vertex displaced from the primary interaction point and an invariant mass within 10 MeV$/c^2$ of the $K^0_S$ mass.  

Fully reconstructed $B$ mesons are 
selected on the basis of 
the beam-constrained mass of the candidate, $M_B = \sqrt{E^2_{beam}-P^2_{reconstructed}}$,
and the difference between the reconstructed and beam energies, $\Delta E = E_{reconstructed} - E_{beam}$. 
$\Delta E$ is sensitive to missing or extra particles in the $B$ candidate, 
as well as incorrect assignment of particle masses. 
For the fully reconstructed $B$ meson decays in this  analysis, 
the $M_B$ distribution peaks at 5.28 GeV$/c^2$ with a resolution ranging between 
2.2-2.6 MeV$/c^2$, 
and $\Delta E$ peaks at zero GeV with a resolution ranging from 16 MeV to 27 MeV.
Candidates are accepted for further analysis if $\Delta E$ 
and $M_B$ are within a signal region 
$\pm 2 \sigma$  around the central signal values for
all channels (except $K^{*-}K^{*+}$ where a larger region is used since this involves two  $K^0_S$'s and is therefore relatively clean).

The backgrounds consist primarily of continuum events 
from $e^+e^- \to q\bar{q}$ ($q=u,d,s,c$) with a 10-15\% contribution
from $B$ decays, and are  
estimated from a combination of off-resonance data and $b \to c$ Monte Carlo. 
Event-shape variables can be used to discriminate against the jet-like continuum 
events since $B$ mesons are produced nearly 
at rest.
Accordingly, we select only events with  $R_2<0.5$, 
where $R_2$ is the ratio of the second to zeroth Fox-Wolfram moments of the 
event\cite{fox}.
In continuum events, momentum conservation aligns the thrust axis of 
the $B$ candidate  
with that of the rest of the event while they are almost uncorrelated  in $B\bar{B}$ events. This allows 
additional suppression of continuum by restricting $|\cos \theta_{tt}|$, 
the angle between the two axes. We require $|\cos \theta_{tt}|<0.7$ for all decay modes, except
for $K^{*+}K^{*-}$, where we use  $|\cos \theta_{tt}|<0.9 $.

The four selection criteria discussed above, on $M_B$, $\Delta E$, 
$R_2$ and $\cos\theta_{tt}$, determine the signal efficiency 
($\epsilon$) for each mode.  We measure this efficiency using Monte
Carlo simulation for
each of the 3 possible helicity states of the decay products: 00, -1-1 
and +1+1. Our study indicates that the 00 helicity has slightly 
lower efficiency than the 11 helicities, 
since it results in more low momentum charged pion and kaon tracks from the 
$B$ decay chain, for which the detector has a lower acceptance. 
In addition, the 00 state will tend to align the vector decay products leading to a higher average $R_2$, also 
decreasing the efficiency. We give separate results assuming the signal 
is 100\% 00 helicity or 100\% $11$ helicity. 
For any assumed helicity distribution of signal events in the data sample, 
upper limits can be obtained by linear interpolation.

We find significant double counting of events in the $K^{*0} \rho^0$ channel, 
caused in most cases by the $K/\pi$ ambiguity in the $K^{*0} \to K^++ \pi^-$ 
sub-decay. In the final results we count only one entry for each event. 
We also consider the possibility of cross-feed between different channels 
of $B \to VV$ decays. Neglecting the contribution from the forbidden decay 
mode $B \to K^{*0} K^{*0}$ ($\Delta S = 2$), the cross-feed effect is small 
even if we use the 90\% upper limits to evaluate the cross-feed contribution 
to the yields.
We do not correct for this contribution when extracting the upper limits.

There are several sources of systematic error. 
A substantial contribution comes from the uncertainty in track efficiency, which is 1.5\% 
per charged track.
For $B$ decay modes with $K^{*\pm}$, there is an additional 5\% uncertainty due to the $K^0_S$ vertex
requirement. 
In addition, we estimate 1\% per charged track uncertainty due to the $dE/dx$ requirement. 
Additional systematic errors include 7\% uncertainty from the thrust criterion and 
3\% from the $\Delta E$ and $M_B$ requirements. 
Uncertainties due to Monte Carlo statistics range from 2\% to 6\%, 
depending on $B$ decay mode. 

\begin{table}[t]
\begin{center}
\caption{
The 90\% C.L. upper limits  for the $B\to VV$ decay modes
$({\cal B}_{\rm CLEO})$ are shown in units of $10^{-6}$,
along with the corresponding theoretical predictions
$({\cal B}_{\rm THEORY})$  \protect\cite{theorists}.
$n_{obs}$ is the number of observed events, $n_{off}$ is the off-resonance background (normalized), 
$n_{b\to c}$ is the $B\bar{B}$ background estimate (from Monte Carlo), and $n_{u.l.}$ is
the corresponding upper limit including systematic error and background statistics. 
The reconstruction efficiency $(\cal E)$ is also 
shown along with the systematic error $(\delta \epsilon)$.
We assume equal branching fractions 
for $\Upsilon(4S)\to B^0 \bar{B}^0$ and $B^+ B^-$. 
}
\begin{tabular}{p{1.5cm}ccccccccc}
 Mode  & Helicity & $\;$$n_{obs}$$\;$ & $n_{off}$ & $n_{b\to c}$ 
       & $\;$$\epsilon$$\;$  & $\delta \epsilon/\epsilon$ & $n_{u.l.}$
       & $\;$${\cal B}_{\rm CLEO}$$\;$ & ${\cal B}_{\rm THEORY}$ \\
       &          &                   &           &      
       &   $(\%)$            &    $(\%)$          &       
       &  ($\times 10^{-6}$) &   ($\times 10^{-6}$)               \\     
\hline \hline
$\rho^0 \rho^0$ & 00 &  54 & 67 & 7.6 & $13$ & 11 &  7.5  & $<$ 5.9 & 0.54--2.5\\
                & 11 &     &    &     & $17$ & 11 &       & $<$ 4.6 &          \\
\hline
$K^{*0} \rho^0$ & 00 &  96 & 92 & 14  & $12$ & 11 &  15   & $<$ 19  & 0.7--6.2\\
                & 11 &     &    &     & $18$ & 11 &       & $<$ 13  &           \\
\hline
$K^{*0} K^{*0}$ & 00 &  22 & 14 & 1.6 & $11$ & 11 &  15   & $<$ 31  &          \\
                & 11 &     &    &     & $14$ & 11 &       & $<$ 24  &          \\
\hline
$K^{*0} \bar{K^{*0}}$ & 
                  00 &  12 & 16 & 1.4 & $12$ & 11 &  5.4  & $<$ 10  & 0.28-0.96     \\
                & 11 &     &    &     & $14$ & 11 &       & $<$ 8.7 &           \\
\hline
$K^{*+} \rho^0$ & 00 &  12 & 5.9  & 2.4 & $ 7.8$ & 13 &  9.5  & $<$ 54 & 0.8--14   \\
                & 11 &     &      &     & $ 12$  & 13 &       & $<$ 36 &           \\
\hline
$K^{*+} \bar{K^{*0}}$ & 
                  00 &   3 & 0.0  & 0.0 & $ 7.3$ & 13 &   5.3 & $<$ 50 & 0.29--1.8 \\
                & 11 &     &      &     & $ 10 $ & 13 &       & $<$ 34 &           \\
\hline
$K^{*+} K^{*-}$ & 00 &   0 & 2.0  & 0.0 & $ 6.6$ & 17 &   2.3 & $<$ 70 &           \\
                & 11 &     &      &     & $ 10 $ & 16 &       & $<$ 45 &           \\
\end{tabular}
\end{center}
\end{table}

The results of this analysis are summarized in Table 1;
we see no statistically compelling signal in any individual decay channel. 
To calculate 90\% confidence level (C.L.) upper limits on the number of signal events in each channel, 
we use the Poisson likelihood of a hypothesis for the average number of signal events $n_S$
given $n_{obs}$ events detected and a background of $n_B=n_{b\to c}+n_{off}$:
\[
{\cal L}(n_S,n_B,n_{obs}) = e^{-(n_S+n_B)} (n_S+n_B )^{n_{obs}} /n_{obs}!
\]
To include the effect of the systematic error in the acceptance and the limited statistics of the background
samples,  we convolute the appropriate distributions (Gaussian for the systematic error, and Poissons for 
the $b\to c$ and continuum backgrounds) with the likelihood function, $\cal L$, to obtain a modified likelihood
function, $\cal L^*$ .
The 90\% C.L. upper limit on the yield is obtained 
by integrating $\cal L^*$, i.e. by solving for $n_{u.l.}$ in:
\[
  \int^{n_{u.l.}}_0{\cal L^*}(n_S,n_B,n_{obs})dn_S = 0.90
\]
The upper limits on the branching ratios are then calculated from the formula,
\[
  {\cal B}(B \to VV) = \frac{n_{u.l.}}{n_{B\bar{B}} \times \epsilon \times \prod_{\rm B}}
\]
where $n_{B\bar{B}}$ is the number of $B\bar{B}$ meson pairs in the data sample, and $\prod_{\rm \cal B}$ is the product over all the relevant branching fractions of the vector meson decay chain.

To summarize, we set 90\% C.L. upper limits on branching fractions of seven $B \to VV$ charmless decay modes. 
Theoretical predictions for the branching fractions of these modes 
tend to be near $10^{-6}$.
Thus our results are consistent with theoretical calculations based on the 
Standard Model. In order to challenge these predictions data samples
of the order of $10^8$ $B\bar{B}$ mesons would be required.


\begin{thebibliography}{1}

\bibitem{soni} D. Atwood and A. Soni, Phys.Rev.D{\bf 59} 13007(1999).

\bibitem{otherCLEO} 
CLEO Collaboration, D. Cronin-Hennessy {\it et al.}, Phys.Rev.Lett.{\bf 85}, 525(2000); \\
CLEO Collaboration, S.J. Richichi {\it et al.}, Phys.Rev.Lett.{\bf 85}, 520(2000); \\
CLEO Collaboration, R. Godang {\it et al.}, Phys.Rev.Lett.{\bf 80}, 3456(1998); \\
CLEO Collaboration, B. Behrens {\it et al.}, Phys.Rev.Lett.{\bf 80}, 3710(1998); \\
CLEO Collaboration, T. Bergfeld {\it et al.}, Phys.Rev.Lett.{\bf 81}, 272(1998); \\
CLEO Collaboration, D.M. Asner {\it et al.}, Phys.Rev.D{\bf 53}, 1039(1996).

\bibitem{theorists} 
A. Ali, G. Kramer,  and C.-D. L\a"u,  Phys.Rev.D{\bf 58}, 94009(1998);\\
A. Deandrea {\it et al.},  Phys.Lett.B{\bf 320}, 170(1994); \\
N. G. Deshpande,  in {\it B Decays},  ed. S. Stone,  World Scientific (Singapore,  1992, 1994); \\
D. Du and Z. Xing,  Phys.Rev.D {\bf 48}, 4155(1993); \\
L. Chau {\it et al.},  Phys.Rev.D{\bf 45}, 3143(1992); \\
G. Kramer and W.F. Palmer,  Phys.Rev.D.{\bf 46}, 2969(1992); \\
M. Bauer,  B. Stech,  and M. Wirbel,  Z.Phys.C{\bf 34}, 103(1987). 

\bibitem{kubota} CLEO Collaboration, Y. Kubota {\it et al.}, Nucl.Instrum.Methods.Res.A{\bf 320},66(1992).
 
\bibitem{fox} G. Fox and S. Wolfram, Phys.Rev.Lett.{\bf 41}, 1581(1978).

\bibitem{pdg} Particle Data Group,  {\it Review of Particle Physics},  
Eur.Phys.Jour.C{\bf 15}, 1(2000). 

\end{thebibliography}
\end{document}